\documentclass[%
 reprint,
 amsmath,amssymb,
 aps,
]{revtex4-1}

\usepackage{graphicx}
\usepackage{dcolumn}
\usepackage{bm}

\begin{document}

\preprint{APS/123-QED}

\title{Observation of Gaussian pseudorapidity distributions for produced\\particles in proton-nucleus collisions at Tevatron energies}

\author{U. U. Abdurakhmanov}
 \email{u.abdurakhmanov@gmail.com}
\affiliation{%
 Institute for Physics and Technology, Fizika-Solntse Research and Production Association,
Uzbek Academy of Sciences, Bodomzor yoli street 2, 100084 Tashkent, Uzbekistan
}
\author{K. G. Gulamov}%
 \email{gulamov@uzsci.net}
\affiliation{%
 Institute for Physics and Technology, Fizika-Solntse Research and Production Association,
Uzbek Academy of Sciences, Bodomzor yoli street 2, 100084 Tashkent, Uzbekistan
}%
\author{V. V. Lugovoi}
 \email{lugovoi@uzsci.net}
\affiliation{%
 Institute for Physics and Technology, Fizika-Solntse Research and Production Association,
Uzbek Academy of Sciences, Bodomzor yoli street 2, 100084 Tashkent, Uzbekistan
}
\author{V. Sh. Navotny}
 \email{navotny@uzsci.net}
\affiliation{%
 Institute for Physics and Technology, Fizika-Solntse Research and Production Association,
Uzbek Academy of Sciences, Bodomzor yoli street 2, 100084 Tashkent, Uzbekistan
}%


\date{\today}

\begin{abstract}
The statistical event-by-event analysis of inelastic interactions of protons  in emulsion at 
$800$ GeV  reveals the existence of group of events with  Gaussian pseudorapidity distributions 
for produced particles, as suggested by hydrodynamic-tube model. Events belong to very central 
collisions of protons with heavy emulsion nuclei with probability of realization of less than 
$1$\% and with multiplicity of shower particles exceeding ($2-3$ times) the average multiplicity 
in proton-nucleus collisions in emulsion. The Bjorken's energy density for these events 
reaches $2.0$ GeV per fm$^3$. The data are interpreted as a result of the QCD  phase transition 
in proton-tube collisions at Tevatron energies. 
\begin{description}
\item[PACS numbers]
\end{description}
\end{abstract}

\pacs{Valid PACS appear here}
\maketitle


Measurements at RHIC and CERN have discovered that the central collisions of heavy ions at these 
energies result initially in production of hadronic matter in the form of very hot compressed 
and a nearly frictionless liquid - quark-gluon plasma (QGP), whose evolution and decay produces 
the final state particles (for reviews see e.g. \cite{ref1,ref2}). The data were analyzed in 
the framework of various theoretical approaches, including different, sometimes very sophisticated, 
versions of the hydrodynamic model \cite{ref3,ref4,ref5}. Recent data from the LHC 
\cite{ref6,ref7,ref8,ref9} reveal unexpected indications on formation of QGP in pp and p$^{208}$Pb 
interactions and provide support to the idea of hydrodynamic model - interaction of an incident 
proton with a tube of nuclear matter. 

The hydrodynamic model of multiparticle production was introduced for the head-on nucleon-nucleon 
collisions  at very high ($>1$ TeV) energies of a projectile \cite{ref10}, and then was 
generalized to the case of nucleon-nucleus collisions \cite{ref11}. In the latter case the 
projectile nucleon can cut out in the nucleus a tube whose cross section is equal to the cross 
section of the nucleon and interacts only with this part of the target nucleus. The length of 
a tube may vary in dependence on the geometry of an interaction.   The projectile and the tube 
undergo a strong Lorentz contraction in the reference frame where their velocities are equal to 
each other. In contrast to the case of a nucleon-nucleon collision, an intricate mechanism of 
compression of nuclear matter treated as a continuous medium comes into play at the first stage 
of collision with a tube. The hadronic matter within the tube has very high density and high 
temperature $T \gg \mu c^2$, where $\mu$ is the pion mass, so that following modern concepts it 
consists of point-like quarks and gluons, rather than usual hadrons. It is the quark-gluon plasma 
and it expands according to the laws of relativistic hydrodynamics of ideal fluid. While expanding, 
it becomes cooler. When the temperature of hadronic matter reaches $T \approx \mu c^2$ , the 
plasma decays producing the final state particles, mostly pions.

The original Landau hydrodynamic model is, to our best knowledge, the only model suggesting some 
certain shape for pseudorapidity distributions of produced particles in both nucleon-nucleon and 
nucleon-nucleus collisions at very high energies - Gaussian distribution. Of course, it is necessary 
to note that only in the case of a very high multiplicity does the pseudorapidity distribution of 
produced particles in an individual event become a meaningful concept.

Simplicity of the model probably is one of the reasons why it is considered to be a "wildly extremal 
proposal" \cite{ref1}. Of course, the scope of the original model was rather narrow. It was introduced 
to describe only few general characteristics of multiparticle production, pseudorapidity distribution 
of charged particles representing one of the simplest characteristics of the production process. 
More advanced versions of the hydrodynamic model are needed to explain more complex characteristics 
and features of the process and from this point of view they are more plausible but maybe less certain 
in predictions.  

Pseudorapidity distributions of particles produced in interactions with nuclei at very high energies 
were discussed in many papers but in most cases the data were presented for inclusive and 
semi-inclusive reactions. At the same time when discussing the shape of pseudorapidity distributions 
of produced particles in the framework of the hydrodynamic model it is better to analyze the 
experimental data on the event-by-event basis  in order, at least, to avoid problems related with the 
geometry of a hadron-nucleus interaction.

In the present paper we are analyzing on the event-by-event basis the shape of pseudorapidity 
distributions of relativistic singly-charged (shower in emulsion terminology) particles (mostly pions) 
produced in inelastic incoherent interactions of $800$ GeV protons with emulsion nuclei. We are
looking at the possibility that these distributions for the individual central collisions are Gaussian 
distributions as suggested by the original hydrodynamic model.

The experimental data of the present study were collected in the framework of the 
Baton-Rouge-Krakow-Moscow-Tashkent Collaboration. We use for the analysis $1800$ inelastic incoherent 
events. 
In each event found, the multiplicity of different types of charged particles were determined and spatial ($\theta$) and azimuthal ($\varphi$) emission angles were measured.

According to the terminology adopted in emulsion experiments, depending on the ionization produced, the charged particles emitted during the interaction were divided into the following groups:

1.\,"$shower$" or $s$-particles  --  singly charged particles with a speed of $\beta \geq 0.7$. These are  mainly particles produced by the interaction of particles (mainly $\pi-$ and $K-$ mesons) and singly charged projectile fragments. Ionization on the tracks of these particles is $I<1.4I_0$, where $I_0$ is the minimal ionization on the tracks of singly charged particles.

2.\,"$gray$" or $g$-particles -- particles moving at a speed of  $\beta < 0.7$ and leaving the tracks with the length of $>3\,mm$ and ionization $I>1.4I_0$ in the emulsion. They  mainly consist of protons knocked out of the target nucleus in the process of interaction and having a momentum of $0.2\leq p \leq 1 \text{ GeV/c}$, with a small admixture of $\pi$-mesons with a momentum of $60 \leq p \leq 170\, \text{ GeV/c}$.

3.\, "$black$" or $b$-particles -- most of them are protons with a momentum of  $ p \leq 0.2 \text{ GeV/c}$ and heavier fragments of the target nucleus, leaving the tracks with the length of $<3\,mm$ and ionization $I>1.4I_0$ in the emulsion.

Details of the experiment together with the main experimental results on multiplicities and 
pseudorapidity distributions were published in \cite{ref12}.

For the analysis of experimental data on the shape of  pseudorapidity distributions of relativistic 
shower particles in individual events we have utilized the statistical approach described in details 
in \cite{ref13}. We use the coefficient of skewness  $g_1$, as a measure of asymmetry, and  the 
coefficient of excess $g_2$, as a measure of flattering, which represent parametrically invariant 
quantities defined as 
\begin{eqnarray}
\label{eq01}
g_{1} &=& m_{3} m_{2}^{-3/2}  \;  , \;\;\;\; \notag \\
g_{2} &=& m_{4} m_{2}^{-2} - 3   \; , \;\;\;\;  \\ 
m_{k} &=& \frac{1}{n} \sum_{i=1}^{n} \left( \eta_{i} - \bar{\eta} \; \right)^{k} \; ,  \;\;\;\;    \bar{\eta} =  \frac{1}{n} \sum_{i=1}^{n} \eta_{i}  \;\;  \notag
\end{eqnarray}
where $m_k$ are the central moments of $\eta$ -distributions and $n = n_s$ stands here for the
multiplicity of $s$-particles in an event.

It follows from the mathematical statistics that if quantities $\eta_{1}$, $\eta_{2}$, ... , $\eta_{n}$ 
are independent of one another in events of a subensemble and obey Gaussian distributions, the distribution 
of these parametrically invariant quantities does not depend on the parameters of
the Gaussian distributions, and the number n of particles in the subensemble event 
uniquely determines the distribution of parametrically invariant quantities. In this 
case the mathematical expectation values and variances of $g_1$ and $g_2$ are 
as follows
\begin{eqnarray}
\label{eq02}
\nu_{g_{1}} (n) &=& 0   , \;\;\;  \sigma_{g_{1}}^{2} (n) =  6 (n-2) (n+1)^{-1} (n+3)^{-1} ,\notag\\ 
\nu_{g_{2}} (n) &=&  - 6 (n+1)^{-1}  , \;\;   \\
\sigma_{g_{2}}^{2}(n) &=&  24 n (n-2) (n-3) (n+1)^{-2} (n+3)^{-1} (n+5)^{-1}.\notag
\end{eqnarray}
We refer to the model described above, where the pseudorapidities obey a Gaussian
distribution, as the $G$ model.

From the mathematical point of view, our goal is to test the hypothesis that
pseudorapidities in the events with different and sufficiently large multiplicity $n$ 
are finite representative random samples with the volume $n$ from the single infinite 
parent population (see Sect.13.3 in \cite{ref13}), in which pseudorapidities are 
distributed according to the Gaussian law. To test this hypothesis, we use the central 
limit theorem (see Sections 17.1-17.4 in \cite{ref13}), which asserts that the sum of 
a large number of independent and equally distributed so-called normalized random 
variables (see Sect.15.6 in \cite{ref13}) has a normal distribution in the limit. In
mathematical statistics, these normalized quantities are constructed from the random variable
and the mathematical expectation and variance obtained from these random variables (see
Sect.15.6 in \cite{ref13}). However, our goal is to test the hypothesis of the normality of 
pseudorapidity distribution in individual experimental events (that is, in the individual 
finite samples from an infinite parent population). Therefore, we construct a normalized 
random variable in a different way, namely: when constructing it for each individual event 
with a multiplicity of $n_s$, we calculate the quantities $g_1$ and $g_2$ (see Eq.~(\ref{eq01})), 
using the experimental values of the event pseudorapidities, and the variances and
mathematical expectations are determined by theoretical formulas (\ref{eq02}) (see Eq.~(29.3.7) 
in \cite{ref13}) for a normally distributed quantity.

Thus, if our hypothesis of normality is true (if the G-model is realized), then by our
construction, the normalized quantities $d_1$ and $d_2$ (see Sect.15.6 in \cite{ref13})
\begin{eqnarray}
\label{eq03}
d_{1} &=& \left[  g_{1} - \nu_{g_{1}} (n) \right] \;  \sigma_{g_{1}}^{-1} (n) \;,  \notag \\
d_{2} &=& \left[  g_{2} - \nu_{g_{2}} (n) \right] \;  \sigma_{g_{2}}^{-1} (n) \;
\end{eqnarray}
have dispersions equal to $1$ and mathematical expectations equal to $0$ both in the subensemble
of events (with the fixed number of particles $n$) and, consequently, in the ensemble of the
events (where $n$ can take any possible values).

Moreover, if the hypothesis of the normality of the pseudorapidtiy distribution is true, then,
according to the central limit theorem of mathematical statistics, for a sufficiently large number
$N$ of independent random samples (that is, the number of interaction events) the sums of these
independent and identically distributed normalized quantities
\begin{eqnarray}
\label{eq04}
\bar{d_{1}} \; \sqrt{N}  =  \frac{1}{\sqrt{N}}  \sum_{i=1}^{N} d_{1i} , \;\; \;\;
\bar{d_{2}} \; \sqrt{N}  =  \frac{1}{\sqrt{N}}  \sum_{i=1}^{N} d_{2i} \; 
\end{eqnarray}
should be less than $2$ with the probability of $95$\% (see Sections 17.1-17.4 in \cite{ref13})

If the hypothesis of normality is true, then the $G$-model is quite realistic and for a small $N$
we can use the asymptotic normality of $g_1$ and $g_2$ in the subensemble of the events described 
by the $G$-model. Then the normalized quantities $d_1$ and $d_2$ are equally
distributed with parameters $0$ and $1$ in both the subensemble and in the ensemble of events
with the large enough $n^{min}$, which is selected  to make the notion of distribution 
meaningful. In this case (for the $G$-model) the sums (\ref{eq04}) have the same restrictions.

In this paper, the sums (\ref{eq04}) were calculated for events with the multiplicity of 
shower particles $n_s$ within the limits from  $n_s^{min}$ to $n_s^{max}$. Calculations
were repeated for different intervals ($n_s^{min}$, $n_s^{max}$) with fixed $n_s^{max}$, 
whereas the value of $n_s^{min}$ was changing from some minimum value of $n_s$  to the maximum 
value of $n_s  = n_s^{max}$, which was defined from the experiment.
\begin{figure}
\centering
\resizebox{0.48\textwidth}{!}{
  \includegraphics{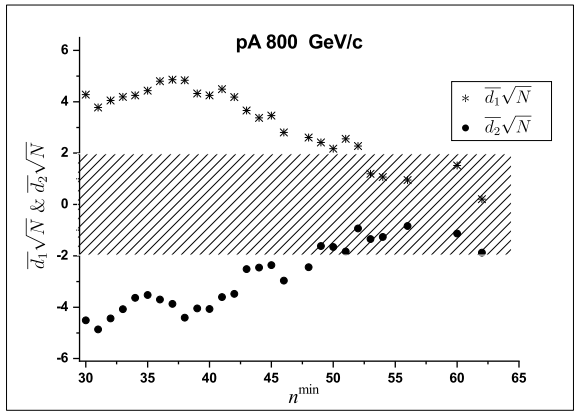}}
\caption{Dependence of parameters $\overline{d_1}\sqrt{N}$ and $\overline{d_2}\sqrt{N}$ on 
$n^{min}$ for proton-nucleus interactions in emulsion at $800$ GeV. The shaded area is the area where $ |\overline{d_1}\sqrt{N}| \; \& \; |\overline{d_2}\sqrt{N}| \; <\; 2$.}
\label{fig:1}       
\end{figure}
\begin{table}
\centering
\caption{Characteristics of $9$ events selected by the statistical approach from 
proton-nucleus  collisions}
\label{tab:1}       
\small
\begin{tabular}{c|c|c|c|c|c|c}
\hline
\hline
\multicolumn{7}{c}{ $p$A $800$ GeV/c}\\
\hline
No &~ $n_s$ ~~&~ $N_h$ ~&~~$\langle \eta \rangle $ ~&~ $\sigma(\eta)$ ~~&~~ $d_1\sqrt{N}$ ~~&~~ $d_2\sqrt{N}$ \\
\hline
1 &  53 &  37 &~~  2.74$\pm$0.35 ~~&  2.58 &  -0.87 &  -.042 \\
2 &  53 &  14 &~~  3.29$\pm$0.16 ~~&  1.15 &   1.62 &  -0.27 \\
3 &  55 &  13 &~~  2.83$\pm$0.17 ~~&  1.23 &  -0.11 &  -1.03 \\
4 &  55 &   9 &~~  2.92$\pm$0.19 ~~&  1.41 &   0.82 &  -0.43 \\
5 &  59 &  19 &~~  3.29$\pm$0.19 ~~&  1.44 &  -0.91 &   0.39 \\
6 &  61 &  11 &~~  2.46$\pm$0.14 ~~&  1.10 &   1.66 &  -0.21 \\
7 &  61 &  19 &~~  3.08$\pm$0.18 ~~&  1.42 &   1.08 &   0.59 \\
8 &  65 &  27 &~~  2.82$\pm$0.17 ~~&  1.40 &   0.53 &  -1.09 \\
9 &  65 &  18 &~~  3.11$\pm$0.14 ~~&  1.12 &   0.24 &  -1.55 \\
\hline
\hline
\end{tabular}
\end{table}
\begin{figure}
\centering
\resizebox{0.48\textwidth}{!}{
  \includegraphics{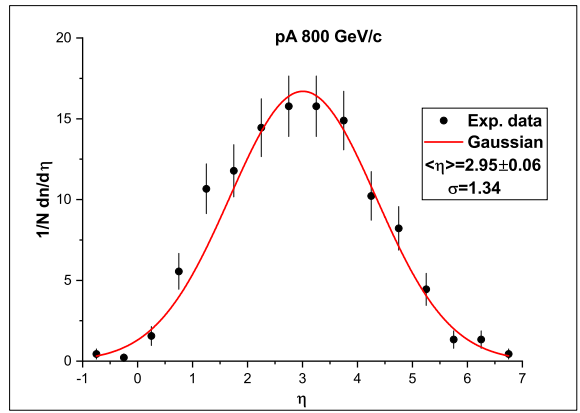}}
\caption{Pseudorapidity distributions of shower particles in selected proton-nucleus 
collisions. Curve is a Gaussian distribution.}
\label{fig:2}       
\end{figure}

The procedure described was applied to the experimental data. In Figure 1 we show dependences 
of the parameters  $\overline{d_1}\sqrt{N}$ and $\overline{d_2}\sqrt{N}$  on the multiplicity 
$n^{min}$ for proton-nucleus interactions considered. It is seen that both parameters  
$\overline{d_1}\sqrt{N}$ and $\overline{d_2}\sqrt{N}$  decrease in their absolute magnitude 
with increasing $n^{min}$. The data reveal the existence in $p$-nucleus interactions of a small 
group of events with the values of parameters $\overline{d_1}\sqrt{N}$ and $\overline{d_2}\sqrt{N}$  
which are simultaneously less than $2$ in their absolute magnitudes. It follows from our 
consideration that pseudorapidity distributions of shower particles in these events are 
representative samplings from the parent Gaussian distribution. Pseudorapidity distributions 
in these individual events obey the Gaussian law. The number of these events in proton-nucleus 
collisions considered is equal to $9$. Characteristics of these selected events are presented 
in Table 1 and in Figure 2 we show the summary pseudorapidity distribution for these events.
\begin{figure}
\centering
\resizebox{0.48\textwidth}{!}{
  \includegraphics{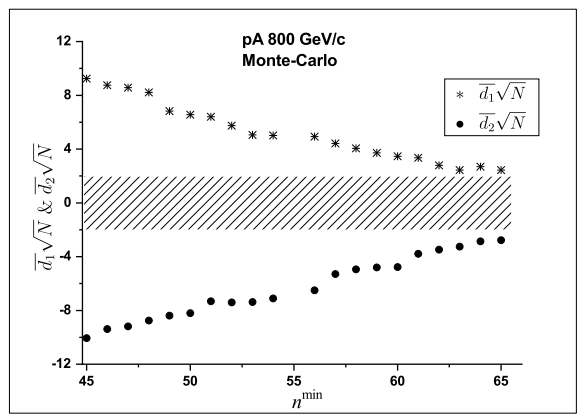}}
\caption{Dependence of parameters $\overline{d_1}\sqrt{N}$ and $\overline{d_2}\sqrt{N}$ on 
$n^{min}$ for $5000$ Monte Carlo generated events. The shaded area is the area where $ |\overline{d_1}\sqrt{N}| \; \& \; |\overline{d_2}\sqrt{N}| \; <\; 2$.}
\label{fig:3}       
\end{figure}

In order to verify our experimental results we have utilized this statistical approach to 
the samples of Monte Carlo events generated following the simple Independent Emission 
Model (IEM) \cite{ref14,ref15}. In the framework of this model we assume that: 
(i) multiplicity ($n_s$) distributions of simulated events reproduce the experimental 
distributions for the interactions considered; (ii) one-particle pseudorapidity distributions 
of $s$-particles in each one of simulated subensembles of events (within, for instance, 
the fixed range of $n_s$) reproduce the experimental distribution for the same range of $n_s$; 
(iii) emission angles of $s$-particles in each one of simulated events are statistically 
independent.

In Figure 3 we show the values of parameters  $\overline{d_1}\sqrt{N}$ and $\overline{d_2}\sqrt{N}$  
in dependence on the multiplicity $n^{min}$ for Monte Carlo events generated in the framework of 
IEM following the experimental multiplicity and pseudorapidity distributions of shower particles 
in proton-nucleus interactions in emulsion at $800$ GeV. We see that absolute values of both 
$\overline{d_1}\sqrt{N}$ and $\overline{d_2}\sqrt{N}$ decrease with increasing $n^{min}$, but 
we found no events with Gaussian  pseudorapidity distributions. We conclude from these results 
that the probability of accidental formation of Gaussian pseudorapidity distributions in individual 
events, not recognizable by the statistical approach utilized, is negligibly small for our 
experimental conditions.

It is necessary to note that the experimental events found by the statistical analysis are very 
rare with probability of realization less than 1\%. They belong to central interactions of hadrons 
with heavy emulsion nuclei. For instance, the average multiplicity of shower particles in these $9$ 
events $n_s = (58.6 \pm 1.6)$ exceeds almost three times the average multiplicity in proton-nucleus 
interactions in emulsion at $800$ GeV, which is equal to  $(20.0 \pm 0.3)$ \cite{ref12}. Average 
multiplicities of black and grey particles, representing, following emulsion terminology, 
fragments of the target nucleus equal respectively for these events  $n_b = (10.0 \pm 1.7)$,  
$n_g = (8.6 + 1.9)$, indicating that central interactions of protons indeed took place with heavy 
(Br,Ag) nuclei in emulsion. Note that $N_h=n_g +n_b$ in Table 1.

It follows from the data on average pseudorapidities and dispersions of pseudorapidity 
distributions in these selected events (see Table 1) that they do fluctuate considerably. Therefore 
the sum of pseudorapidity distributions in selected individual events shown in Figure 2 do not 
demonstrate very good agreement with the Gaussian shape.  From Figure 2  for the density 
\scalebox{1.14}{$\frac{1}{N} \frac{dN}{d\eta}$} in the central region for selected proton-nucleus 
collisions we have $(15.9 \pm 1.8)$.          

The experimental observations of the present paper encourage us to interpret the existence of 
events with the Gaussian pseudorapidity distributions of produced particles in central relativistic 
proton-nucleus interactions as a result of a proton-tube collisions and subsequent  formation in the 
course of an interaction of a droplet of hadronic matter - the quark-gluon plasma, i.e. the 
primordial high density state, whose expansion and cooling leads to its decay with production of 
final state particles. This interpretation may be supported by following considerations. 

Calculations in the framework of lattice QCD show \cite{ref16, ref17} that at the energy densities 
exceeding a critical value of about $1$ to $1.5$ GeV per fm$^3$, achievable at incident energies 
of about  $\sqrt{s_{NN}} \gtrsim 5\text{ GeV} $, the hadronic phase of matter disappears giving 
 rise to the primordial high density state (QGP) whose evolution is governed by the elementary 
interactions of quarks and gluons. From the radius of a tube equal $1$ fm and the experimental 
value of the density \scalebox{1.14}{$\frac{1}{N} \frac{dN}{d\eta}$} for selected proton-nucleus 
events we have for the Bjorken's energy density \cite{ref18} approximately $2.0$ GeV per fm$^3$ 
what is more than the critical value of the density. 

It is known from simple kinematics that the rapidity of the center of mass frame in a proton-tube 
collision, where a tube consists of $k$ nucleons, is shifted from that of a proton-proton collision 
on the value $\Delta y = \frac{1}{2} ln (k)$. So, from the value of this shift it is possible to 
estimate  the average number  of nucleons in the tube. From the experimental data of Figure 2 for 
selected proton-nucleus collisions we have an estimate $k = 4.7$, which leads to the corresponding 
estimate of the energy of proton-tube interactions $\sqrt{s} \sim 80\text{ GeV} $. At the same time 
the Glauber model gives for the average number  of intranuclear collisions $2.75$ and $3.20$ for 
$p$-Em and $p$-BrAg interactions, respectively \cite{ref19, ref20}.

Of course, realization of the phase transition cannot be easily expected in proton-nucleus 
collisions at these energies, even central ones. If it nevertheless does happen it must be a rare 
and random phenomenon with fluctuations and instabilities playing significant role in outcome of 
an interaction, so that the produced intermediate QCD objects may vary in some important initial 
characteristics, in the volume, for example. Similar situation was considered for heavy-ion 
collisions at SPS energies \cite{ref21}. It was shown that big droplets of quark-matter may be 
formed at this energy densities due to fluctuations, but not in average events. The percolation 
model was used to reflect the complexity of the process.  Therefore it was recommended to search 
for these objects on the event-by-event basis. Evolution of these objects follows obviously general 
principles taken into account by the original hydrodynamic model and may lead to the Gaussian 
distributions of final state particles in pseudorapidities. Therefore we believe that it is 
important to study and to confirm this possibility in other experiments as well. 

We are grateful to all members of BRKMT-Collaboration for the joy to work together and for 
the excellent quality of the data.

\bibliography{paper}

\end{document}